\documentclass[aps,prd,a4paper,superscriptaddress,nofootinbib,
showpacs,showkeys,amsfonts,amssymb,amsmath]{revtex4}
\usepackage{amssymb,latexsym}
\usepackage{amsmath, amsthm}
\usepackage{amscd}
\usepackage{times}
\usepackage{epsfig}
\usepackage{graphicx}

\begin{document}
\title[Equilibrium configurations]{Equilibrium configurations from
gravitational collapse}
\author{Pankaj S. Joshi} \email{psj@tifr.res.in}
\affiliation{Tata Institute of Fundamental Research, Homi Bhabha Road,
Colaba, Mumbai 400005, India}
\author{Daniele Malafarina} \email{daniele.malafarina@polimi.it}
\affiliation{Tata Institute of Fundamental Research, Homi Bhabha Road,
Colaba, Mumbai 400005, India}
\author{Ramesh Narayan} \email{rnarayan@cfa.harvard.edu}
\affiliation{Harvard-Smithsonian Center for Astrophysics,
60 Garden Street, Cambridge, MA 02138, USA}
\swapnumbers

\begin{abstract}
We develop here a new procedure within Einstein's theory of gravity to
generate equilibrium configurations that result as the final state of
gravitational collapse from regular initial conditions. As a
simplification, we assume that the collapsing fluid is supported only
by tangential pressure. We show that the equilibrium geometries
generated by this method form a subset of static solutions to the
Einstein equations, and that they can either be regular or develop a
naked singularity at the center. When a singularity is present, there
are key differences in the properties of stable circular orbits
relative to those around a Schwarzschild black hole with the same
mass. Therefore, if an accretion disk is present around such a naked
singularity it could be observationally distinguished from a disk
around a black hole.

\end{abstract}
\maketitle

\section{Introduction}

It is well known that any spherical distribution of non-interacting
particles (thus in the absence of pressure), cannot sustain itself
against the pull of its own gravitational field and must undergo
complete gravitational collapse. The final outcome of such a dust
collapse is either a black hole or a naked singularity, depending on
the initial density configuration and velocity distribution
of the particles
\cite{dust}.
In recent years, from the study of gravitational collapse in a wide
variety of scenarios --- self-similar collapse
\cite{selfsim},
scalar fields
\cite{scalar},
perfect fluids
\cite{perfectfluid}
and other general forms of matter fields --- it has
emerged that black holes and naked singularities have both to
be considered as possible final outcomes of complete collapse
of a massive matter cloud in general relativity 
\cite{press}.
This is true also for alternative theories of gravity
such as $f(R)$ gravity and Lovelock gravity
\cite{alt}.

When non-vanishing pressures are present within the matter cloud
-- a physically more realistic scenario than dust -- complete gravitational
collapse is not the only possible final state.
Hence, a question of much physical interest is, if and under what conditions
can we have equilibrium configurations that originate from dynamical
gravitational collapse within general relativity.
In fact, we know that
equilibrium non-singular gravitating systems exist in nature,
e.g., planets, stars, galaxies, all of which
form via gravitational collapse, and that these objects
are of great significance from the perspective of
both theory and observations.

We show here that the gravitational collapse of a matter cloud with
non-vanishing tangential pressure, from regular initial data, can give
rise to a variety of equilibrium configurations as the collapse final
state. We consider here, for the sake of simplicity and clarity, the
case of collapse with non-zero tangential pressure and vanishing
radial pressure.  While such a simplified model
represents in some sense an idealized example, the case of a more
realistic matter source, for example composed of a perfect fluid
(with both radial and tangential pressure),
can also be investigated and its equilibrium configurations will be
  reported separately in a future work.
Collapse with only tangential pressure
  provides a relatively transparent structure (mainly because the mass
  function is conserved as we explain below), and
  this case is of much interest in its own right,
  since it reveals a rich array of possibilities. Therefore it is important and
  useful to study and understand these simpler models in order to figure out
  the possibilities that they offer on the collapse final states.

Models with vanishing radial pressure were first
investigated by Datta in the special case where the
matter cloud is composed of counter-rotating particles in the so
called Einstein cluster geometry \cite{cluster}.  Over the past
years, these models have been studied in detail in order to
characterize the final outcome of complete gravitational collapse in
terms of black holes and naked singularities \cite{tangential}.  We
use here the general formalism developed in \cite{IJMPD} to study the
dynamical evolution of a massive matter cloud that collapses from
regular initial data. We develop a technique and procedure that
allows us to investigate when such a collapse can halt to
form an equilibrium configuration. We show
that from such a collapse process static
configurations can arise which are either regular or singular
at the center. We then investigate a particular toy model belonging
to this class which presents a central naked singularity and
we study the physical properties of an accretion disk in this spacetime.

The main noteworthy feature of the work
presented here is that
it shows the existence of solutions in which naked singularities
arise from a dynamical process of gravitational collapse, starting
from regular initial conditions.
This makes the models useful and interesting from a physical
point of view since many other naked singular geometries that
are usually investigated within general relativity are
not obtained  from the dynamical evolution of a collapsing
cloud with regular initial data (although we note that there exist
some much discussed examples of ways to overspin a Kerr black hole,
thus turning it into a naked singularity by dropping spinning
extended bodies through the horizon, see for example
\cite{deF}).
We point out here that although we deal with an
idealized toy model, where the singularity is approached
asymptotically, physically it can be interpreted as a slowly
evolving collapse scenario where the central density is increasingly
high (to a point where quantum effects would become relevant),
and which is always visible to faraway observers.
At later and later times, as
the gravitational collapse evolves, the collapse model will
be arbitrarily close to the idealized final static configuration, with the
rate of collapse becoming arbitrarily small. Therefore
such a `freezing' of the dynamics as $t$ grows
allows us to neglect the collapse at a late
enough time when the central region has achieved an extremely
high density.  At this late time, the approximation to a static configuration
holds to a high degree of accuracy, and it is meaningful to
study its physical properties, e.g., the nature of an accretion disk
in the static configuration.

The properties of circular orbits
for a non-rotating source sustained only by tangential pressure, as
discussed here, differ considerably from corresponding results
evaluated for the Kerr spacetime (see for example
\cite{Kerr}).
Furthermore, the objects we study will have different optical
properties in terms of shadows and gravitational lensing compared to
well studied analogues
in other static and stationary geometries.
In fact, it has been shown that whenever a photon sphere is
present, as is the case in certain Kerr geometries, the effects of
gravitational lensing make a naked singularity
indistinguishable from a black hole
\cite{Virb}.
In the model presented here, we find naked singularities with no photon sphere,
thus leaving open the possibility to distinguish such objects
from a black hole.

While the arguments presented above provide a number of motivations
for the present study, the main objective of this work
is to show the existence of models in which asymptotically static
spacetimes with and without naked singularities arise as a result of
dynamical gravitational collapse from regular initial
configurations.  It is briefly shown that
observational features e.g., of an accretion disk,
in such models can be quite different
from those of a Schwarzschild black hole, though a more
detailed investigation of the properties of accretion disks,
including a parameter study of energy flux and luminosity,
is deferred to a future work.

In section \ref{collapse}, we give a brief overview of
the general formalism to describe the dynamical evolution of
a matter cloud sustained by tangential pressure. This formalism
can be applied to the
case of collapse, expansion, bounce or asymptotic equilibrium. In
section \ref{equilibrium} we impose the condition that the system
asymptotes to a state of equilibrium, and we then analyze under
what circumstances a collapsing cloud can settle to such a
static limit. Section \ref{applications} is then devoted to the
analysis of a toy model for which the stable circular orbits
and related physical properties of interest are presented.
This is used to establish when, and in what manner, such models
can be observationally distinguished from a black hole.
Finally, in section \ref{remarks} we highlight the main results
and indicate perspectives for future research.

\section{Gravitational collapse} \label{collapse}

The spherically symmetric spacetime metric describing a dynamical
gravitational collapse can be written as,
\begin{equation}\label{metric}
    ds^2=-e^{2\nu}dt^2+\frac{R'^2}{G}dr^2+R^2d\Omega^2 \; ,
\end{equation}
where $\nu$, $R$ and $G$ are functions of the comoving time $t$ and
the comoving (Lagrangian) radial coordinate $r$.  In the case of
vanishing radial pressure the energy-momentum tensor is given by
$T^0_0=\rho, \; T_1^1=0, \; T_2^2=T_3^3=p_\theta$, and the Einstein
equations take the form
\begin{eqnarray}\label{pr}
  p_r &=&-\frac{\dot{F}}{R^2\dot{R}}=0 \; , \\ \label{rho}
  \rho&=&\frac{F'}{R^2R'} \; , \\ \label{pt}
  p_\theta &=&\frac{1}{2}\rho R \frac{\nu'}{R'} \; ,\\ \label{Gdot}
  \dot{G}&=&2\frac{\nu'}{R'}\dot{R}G \; .
\end{eqnarray}
In the above, $F$ is the Misner-Sharp mass, which describes the amount
of matter enclosed by the shell labeled by $r$, and is given by
\begin{equation}\label{misner}
F=R(1-G+e^{-2\nu}\dot{R}^2) \; .
\end{equation}
Equation \eqref{pr}, which results from the assumption of
pure tangential pressure,
immediately implies that $F=F(r)$, and
so the mass
interior to any Lagrangian radius $r$ is conserved throughout the
evolution.
Therefore, at all times the metric describing the evolving
cloud can be matched to an exterior Schwarzschild solution with a
total mass $\mathbf{M_{TOT}}$ at a boundary $r=r_b$, which corresponds
to a time-dependent physical radius $R_b(t)=R(r_b,t)$ \cite{matching}.

We note that it is also possible to consider a
more general collapse problem where we allow for non-zero radial
pressure. In that case, the considerable simplification of being
able to match to Schwarzschild may no longer be possible. However,
matching to a generalized Vaidya spacetime is always possible
for any general type I matter field as also for fields with non-zero
radial pressure, subject to satisfying energy conditions and
other reasonable physical regularity conditions. We shall
consider such a more general situation separately, but we focus here
on the simpler case of zero radial
pressures since it offers considerable clarity and transparency
on the role that pressure plays in the dynamical evolution of
gravitational collapse. This also allows us to see clearly how
the presence of pressure permits equilibrium solutions, as opposed
to the case of dust where such solutions are never possible.
Models with non-zero tangential pressure have been of much
interest in the past as mentioned above, though this interest
was mainly focused on finding the collapse final states in terms
of black holes or naked singularities.

There is a scaling degree of freedom available in the
definition of the physical radius $R$. Using this we introduce
a scaling function $v(r,t)$ defined by
\begin{equation}
R(r,t)=r\,v(r,t), \qquad v(r,t_i)=1,
\end{equation}
where the latter condition simply states that $R=r$ at the
initial time $t=t_i$ from which the collapse develops.
In the following we assume that no shell-crossing singularities,
defined by $R'=0$, occur during the evolution. These are
supposed to be weak singularities that arise due to
the collision of different radial shells and would be
removable by a suitable change of coordinates. To avoid
shell-crossing singularities we must impose that $R'>0$
during the evolution. This in turn implies that the weak
energy conditions are satisfied for positive pressures
during collapse whenever $F'>0$.\\

Our main procedure now for evolving the gravitational collapse
is as follows. We have six unknowns, namely $\rho$,
$p_\theta$, $\nu$, $G$, $F$ and $R$, and four Einstein equations,
so we have the freedom to choose two free functions.  Once we
specify the initial data for the above six functions at an initial
epoch $t=t_i$ and specify the two free functions, the system is
closed and the Einstein equations then evolve the collapse to
any future time. Typically, we may choose the free functions
to be the mass function $F(r)$, which specifies the initial mass
which is conserved for the cloud (from which the energy
density can be obtained from equation \eqref{rho}), and the
tangential pressure $p_\theta$. Then, given the initial values, the future
evolution is fully determined by the Einstein equations
\cite{initial}.  We note that, while in the present case the mass
function is time independent and is chosen once and for all, the
pressure depends on $r$ and $t$ via $v(r,t)$ as $p_\theta= p_\theta
(r,v)$. Hence a global choice for the pressure at all times has to be
supplied, which then fully fixes the evolution of the system.  One
could, for example, begin by choosing $F=M_0r^3$, which corresponds to
a matter cloud that is initially perfectly homogeneous (constant
density). It is then known that, for certain classes of choices
of the evolution of $p_\theta$, a complete gravitational collapse
would terminate in either a black hole or a naked singularity
final state \cite{tangential}.  On the other hand, for certain
other choices of the pressure, a bouncing behaviour for the cloud
may result, where the initial collapse is reversed to turn
into an expansion
\cite{bounce}.
It is thus the choice we make for $p_\theta$ that determines which
way the collapse develops and evolves in future: Whether it will
be a continual collapse to either a black hole or a naked singularity,
or there will be a bounce at some stage, or it will settle into
an equilibrium final configuration as we discuss here.

We note that actually any choice of $p_\theta$ is equivalent to a
choice of a constitutive equation of state for the matter, and
vice-versa.  In fact, the relation between the energy density and the
tangential pressure is given implicitly by equation \eqref{pt},
\begin{equation}
   k(r,v)\equiv \frac{p_\theta}{\rho} =\frac{1}{2}R\frac{\nu'}{R'},
\end{equation}
and so $p_\theta$ in general need not have an idealized form such as a
linear or polytropic function of $\rho$.  In fact, it is reasonable to
suppose that the collapsing system will go through very different
regimes, moving from its initial low density and weak gravitational
field configuration to later stages where the density and
gravitational fields might be extremely high. This would be
reflected in the constitutive equation being, in general, a
function of $r$ and $t$, taking into account how the matter
content changes during the evolution.

From equations \eqref{pt} and \eqref{Gdot}, we can solve
for $\nu$ and $G$ to obtain
\begin{eqnarray}
    \nu(r,t)&=&2\int_0^rk\frac{R'}{R}d\tilde{r}+y(t) \; ,\\ \label{G}
    G(r,t)&=&b(r)e^{4\int^1_v\frac{k}{\tilde{v}}d\tilde{v}} \; .
\end{eqnarray}
We ignore the function $y(t)$, which comes from integrating equation
\eqref{pt}, since it can be absorbed in a redefinition of the time
coordinate $t$.  The free function $b(r)$, which comes from
integrating equation \eqref{Gdot}, is related to the velocity
profile of the particles (it is easy to check that in the
pressureless case we get $G=b$). The only unknown that remains
finally is the metric function $R$, which is the physical radius
for the matter cloud and which is given by the solution
of the differential equation for $\dot{R}$ provided by equation
\eqref{misner}.  The interior spacetime metric during
collapse can then be written as
\begin{equation}\label{metric2}
    ds^2= -e^{4\int_0^rk\frac{R'}{R}d\tilde{r}}dt^2+
\frac{R'^2}{b(r)e^{4\int^1_v\frac{k}{\tilde{v}}d\tilde{v}}}dr^2+R^2d\Omega^2 \;.
\end{equation}

In the present case with the spacetime metric given
as above, while certain explicitly
solved interior models such as the Einstein cluster are available,
we note that in general obtaining explicit solutions in fully
integrated form for the Einstein equations is rather difficult except in
the simplest cases. It is also not always necessary.
What we really need here for our purpose is, assuming
regular behaviour of the free functions and the initial data,
information about the structure of the dynamical collapse.
In fact, the behaviour of a collapse solution can often
be inferred in many cases without having to explicitly
carry out the full integrations.

\section{Equilibrium configurations}\label{equilibrium}

We investigate the following question here: For what
choices or classes of the pressure $p_\theta(r,v)$ (or equivalently
$p_\theta(r,t)$), does a cloud that collapses from regular initial
conditions approach asymptotically an equilibrium configuration
with a static spacetime geometry?
In the following, given the freedom to choose the
tangential pressure function $p_\theta$, we construct classes
of models in which the pressure balances the attraction of
gravity in the asymptotic final state, thereby obtaining
final equilibrium configurations. This is subject to
physical conditions such as the positivity of energy
density and the regularity of the initial data.

For any fixed $r$, the equation of motion \eqref{misner}
can be written in terms of $v$
in the form of the following effective potential,
\begin{equation}\label{vdot}
    V(r,v)=-\dot{v}^2=-e^{2\nu}\left(\frac{M}{v}+\frac{G-1}{r^2}\right) \; ,
\end{equation}
where the function $M(r)$ is defined by $F=r^3M(r)$, and a suitable
choice of this function implies that the Misner-Sharp mass is
well-behaved at the center of the cloud ({\it i.e.},  non-singular and
without cusps). In the Newtonian analogy, the negative of the
effective potential describes the weighted kinetic energy of the
particles in the shell labeled by $r$.  Notice that, since
$F$ does not depend on $t$, it must be the same throughout
the collapse and in the final equilibrium state.

From the above equations it is clear that any static
configuration or bounce cannot at all arise in pressureless
dust collapse. This is seen immediately from the fact that
for dust $\dot{v}^2=M(r)/v+f(r)$.  Therefore, if
$\dot{v}<0$ at some time, {\it i.e.}, once collapse has begun,
$\dot{v}<0$ continues to be the satisfied
at all later times.  Further, if the density is not zero,
then $\ddot{v}=0$ cannot be achieved at any later time.
The continual dust collapse thus inevitably terminates
in a spacetime singularity as the collapse final state, which
is either covered within an event horizon thus forming a
black hole, or is a naked singularity. This
final state either way is determined by the nature of the
initial data from which the collapse evolves. The addition
of non-zero pressures changes the situation. In particular,
it also introduces a degree of freedom that can be used
to balance the gravitational attraction to obtain a static
final equilibrium configuration, as we show below.

In order for the system to reach a static configuration
where collapse stabilizes, we require both the velocity and
the acceleration of the in-falling shells to go to a
vanishing value as the collapse progresses in future.
We therefore need the limiting conditions,
\begin{equation}\label{static}
    \dot{v}=\ddot{v}=0,
\end{equation}
which are equivalent to $V=V_{,v}=0$, where
\begin{equation}
    V_{,v}=e^{2\nu}\left(\frac{M}{v^2}-\frac{G_{,v}}{r^2}\right)
    -2\nu_{,v}e^{2\nu}\left(\frac{M}{v}+\frac{G-1}{r^2}\right) \; .
\end{equation}
As the collapse progresses in time, the evolution metric
function $v(t,r)$ approaches in the limit an equilibrium value
$v=v_e (r)$.  The effective potential can be considered as
a function of $v$ for any fixed shell $r$, so from the
conditions (\ref{static}) we obtain the limiting
equilibrium configuration $v_e(r)$.

We note that in the comoving coordinates that we
have used here, the final equilibrium, {\it i.e.}, the
static limit, is reached in the limit of the comoving
time $t$ going to infinity. If we linearize the effective
potential $V$ near the equilibrium configuration $v_e$,
we can write it as $V(v)=H(r)^2(v-v_e)^2$ for a certain
function $H<0$ that can be obtained from the second
derivative of $V$. Therefore we get $dv/(v-v_e)=H(r)dt$,
or $(v-v_e)=\exp[H(r)(t-t_i)]$, which implies
$v\rightarrow v_e$ as $t\rightarrow +\infty$.
It is possible that there might exist some reparametrization
of $t$ that allows the singularity to appear in a finite time. But
it is not necessary to find such a coordinate change explicitly
as we prefer to work with comoving coordinates, which have
a clear physical interpretation and appeal.

Once $M(r)$, or equivalently the mass function $F(r)$,
is chosen, by imposing the conditions \eqref{static} we
obtain two equations that fix the behaviour of $G$ and $G_{,v}$ at
equilibrium in terms of the equilibrium solution $v_e(r)$,
\begin{eqnarray}\label{ve}
  G_{e}(r)&=&G(r, v_e(r))=1-\frac{r^2M(r)}{v_e(r)} \; , \\ \label{M}
  (G_{,v})_e &=&G_{,v}(r, v_e(r)) = \frac{M(r)r^2}{v_e^2} \; ,
\end{eqnarray}
where the velocity profile $b(r)$ that appears in equation
\eqref{G} has been absorbed into $G_e(r)$.
From equation \eqref{rho} evaluated at equilibrium,
we obtain the energy density at equilibrium,
\begin{equation}
\rho_e(r)=\frac{3M+rM'}{v_e^2(v_e+rv_e')},
\end{equation}
while the tangential pressure can be written as
\begin{equation}
p_{\theta e}= \frac{1}{4}\rho_e v_e \frac{(G_{,v})_e}{G_e}
=\frac{1}{4}\frac{r^2M(3M+rM')}{v_e^2(v_e+rv_e')(v_e-r^2M)} \; .
\end{equation}

In any gravitational
collapse that begins from regular initial conditions, {\it i.e.},
that has a regular mass function $F(r)$, and evolves towards
a final equilibrium state, we have the freedom to choose the
final configuration via the function $v_e(r)$. Once we choose
$F(r)$ and $v_e(r)$, these two functions fully
determine all other quantities in the final equilibrium: $\rho_e(r),
G_e(r), (G,_v)_e(r)$ and $p_{\theta e}(r)$.  While this fully
specifies the final state, we still have the freedom to choose
how the system evolves between the initial and final configurations.
As stated earlier, the collapse evolution is fully fixed by
choosing $F(r)$ and $p_\theta(r,v)$, one of which, namely the mass
function $F(r)$, we have already chosen. Thus, in order to go
to the desired final equilibrium configuration as collapse limit,
the class of allowed pressures $p_\theta(r,v)$ is to be so
chosen that we have $p_\theta(r,v) \to p_{\theta e}(r)$ as
$t\to\infty$, where the equilibrium pressure $p_{\theta e}(r)$
was determined by the choice of the free function $v_e(r)$, as
indicated above.  For this entire class of $p_\theta$ evolutions,
the dynamical gravitational collapse will necessarily go in
the asymptotic limit to an equilibrium spacetime geometry,
which is defined by $F(r)$ and $v_e(r)$.

Thus, we can finally write the metric \eqref{metric2}
at equilibrium as,
\begin{equation}\label{metric-eq}
ds^2_e=-e^{4\int_0^r\frac{p_{\theta e} R_e'}{\rho_e R_e}d\tilde{r}}dt^2+
\frac{R_e'^2(\rho_e+4p_{\theta e})}{\rho_e}dr^2+R_e^2d\Omega^2 \; ,
\end{equation}
where $R_e(r)=rv_e(r)$ is the physical radius of the Lagrangian shell
$r$ in the final equilibrium configuration.  A dynamical solution of
the Einstein field equations, as represented by the metric
\eqref{metric2} with $F(r)$ fixed, and with the class of pressures
$p_\theta(r,t)$ chosen such that $p_\theta \rightarrow p_{\theta e}$
(or equivalently choosing a function $v(r,t)$ such that $v \rightarrow
v_e$), will therefore tend to a final equilibrium metric of
the form \eqref{metric-eq}, where in the final state all
the functions depend only on $r$.

In principle, the above equilibrium metric, which is
the final state of collapse, is not required to be necessarily
regular at the center. In fact we can see from equation
\eqref{rho} that since $M$ is finite at $r=0$, whenever
we have $v_e(0)=0$, the energy density at equilibrium diverges
at $r=0$ and the equilibrium metric presents a central
spacetime singularity. Clearly, this singularity has been
achieved as the result of collapse from regular initial data
that respect the energy conditions and it is
somehow similar to those arising in complete gravitational
collapse, though in this case the outer shells do not
fall into the singularity but halt at a finite radius,
thus creating a static compact object.

\begin{figure}[hh]
\includegraphics[scale=0.8]{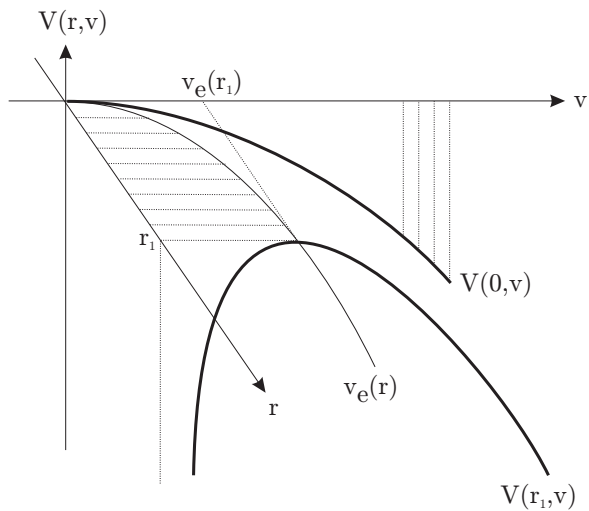}
\caption{The conditions for equilibrium given in equation
\eqref{static} require that, for each $r$, the effective potential
$V(r,v)$ as a function of $v$ should have a maximum, and further
that the value of the potential at the maximum must be zero. Then,
$V(r,v)=0$ implicitly defines $v_e(r)$.}
\label{fig}
\end{figure}

We see from the above that in order to fix completely
the behaviour of $p_\theta$ at equilibrium, we need to give
the explicit form at equilibrium of $G$ and $G_{,v}$ and
both the equilibrium conditions, namely $\dot{v}=0$
and $\ddot{v}=0$, are necessary as well as sufficient for
such a purpose.  The freedom to choose the tangential
pressure as above in the evolving collapse phase allows for
different effective potentials to have different equilibrium
configurations. On the other hand, we can reverse the
reasoning and choose a certain equilibrium configuration
$v_e(r)$, which implies a specific $p_{\theta e}$.  We
can then select the class of tangential pressures with
proper limit such that they give rise to that specific
effective potential (see Fig. \ref{fig}).

If an equilibrium configuration with a singularity at
the center is achieved as the limit of gravitational collapse,
it is important to check under what conditions the singularity
will not be covered by an event horizon. In the static
limit, this simply means that the boundary where the cloud
matches the vacuum Schwarzschild exterior has a radius
greater than the Schwarzschild radius. Therefore, from the
boundary condition at equilibrium we obtain a condition for
the absence of trapped surfaces at equilibrium. In order for
the region near the center to be not trapped, we must have
$F/R<1$. In the final equilibrium state, writing
$G$ in terms of energy density and pressure, we have
\begin{equation}
    \frac{F}{R_e}=1-G_e=\frac{4p_{\theta e}}{\rho_e+4p_{\theta e}} \; ,
\end{equation}
which is obviously smaller than unity for positive energy density and
positive $p_{\theta e}$.  Thus, we obtain the interesting result that
any central singularity that forms in the final equilibrium
configuration via the collapse dynamics described in this paper is
always a naked singularity.  This is not surprising since we are
describing dynamical evolution from regular initial data without
trapped surfaces.  This means that at the initial time $t_i$ the
center of the cloud is not trapped. Since the evolution remains
regular, with the singularity arising as $t$ goes to the asymptotic
limit as the system reaches equilibrium, we can, in principle,
choose any time slice to be the initial time. This implies
that during the whole evolution the center of the cloud is not
trapped and so the central singularity that appears in the
final equilibrium is naked.

It is relevant to mention here that static interior solutions
of the Schwarzschild metric have been studied in the past and
static interiors supported by only tangential pressures were given by
Florides \cite{florides}, who wrote the most general metric in this
case and found solutions depending on the choice of the free function
$F(r)$. A static spherically symmetric line element depends only
on the physical radius $R$, and it is easy to verify that it can be
obtained from the metric \eqref{metric-eq}, once a suitable change of
coordinates $R=R(r)$ is made. It therefore follows that the class of
equilibrium configurations we have obtained here belongs to the family of
static metrics given by Florides.  The Einstein equations are easily
rewritten in this case.  The metric functions $G$ and $\nu$
in the static tangential pressure case are then given by
\begin{equation}\label{stat}
G(R)=1-\frac{F(R)}{R}, \qquad 2\nu_{,R} = \frac{F(R)}{R^2G},
\end{equation}
and equation \eqref{pt} becomes a definition for
the tangential pressure.

The important point is,
the entire family of static tangential pressure solutions
described by Florides can be obtained from gravitational
collapse as the asymptotic equilibrium limit of collapse,
by using the procedure we outlined here.

\begin{figure}[hh]
\includegraphics[scale=0.8]{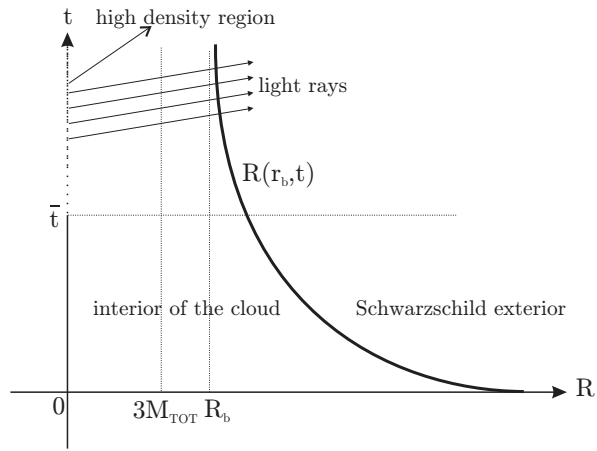}
\caption{The collapsing cloud approaches asymptotically the
equilibrium configuration. For $t>\bar{t}$ the central density
grows to arbitrarily high values. No trapped surfaces are
present at any time. Light rays would escape the strong gravity
region, reach the boundary and then propagate to the exterior.
As $t$ grows the collapsing cloud approximates more and more
the static model. The boundary remains larger than the photon
sphere for the Schwarzchild spacetime at all times.}
\label{fig2}
\end{figure}

The physical relevance of such equilibrium
configurations comes from the fact that in the case of a singularity
at the center, as the comoving time $t$
increases, the collapse slows down and the central density increases
arbitrarily high. For sufficiently large values of $t$ the static
models do therefore approximate the collapsing cloud to an
arbitrarily high degree of accuracy
(see Fig. \ref{fig2}).

As noted earlier, the naked singularity
here, when present in the final equilibrium configuration,
is achieved only asymptotically by the collapsing cloud as
the comoving time goes to infinity.  In the collapsing matter
cloud that approaches the equilibrium, there is no singularity
at the center at any finite time $t$. Even when the divergence
of the energy density occurs at an infinite comoving time,
what is important to note is that the ultra-high density region
that develops at the center of the collapsing cloud continues
to be always visible and never trapped. This is the region where
classical relativity may eventually break down at high enough
densities and quantum effects might occur and dominate.
This phase is always obtained in a finite but large enough
time, before the actual singularity of the equilibrium.
It is the visibility of such a region during collapse, which
approaches in the limit the equilibrium model with an actual
naked singularity at the center, which is the main reason
for our study of the physical properties of such objects. In
contrast, we know that in the case of collapse to a black hole,
the ultra-high density regions are always necessarily hidden
inside the event horizon after a certain stage in the collapse, and
any Planck scale physics that might occur close to the singularity
is invisible to distant observers.

The basic point we make here is: At later and later
times, as the collapse progresses, the interior metric of the
collapsing matter cloud becomes arbitrarily close to that of the
static configuration metric. This is the sense in which the
static or equilibrium configuration is approached
to a higher and higher degree of accuracy as the dynamical collapse proceeds.
Thus the collapse eventually freezes as time evolves, with the
velocities and acceleration of the collapsing shells becoming
arbitrarily small, approaching a vanishing value. The dynamical
collapse asymptotes to the static tangential pressure model to
arbitrary precision at late times, and so the static
model is the limiting configuration to which it converges.

We have thus obtained
here a wide family of gravitational collapse solutions that lead to
equilibrium configurations in the final state. This family of solutions
has the following degrees of freedom: Firstly, we have the
freedom to choose any form of the equilibrium function $v_e(r)$. Then,
corresponding to that $v_e(r)$, we have a wide family of tangential
pressures $p_\theta(r,v)$, or equivalently $p_\theta(r,t)$, which we
can choose from, all of which give the same final equilibrium
state as a result of dynamical collapse.

\section{Physical applications}\label{applications}

In this section, we study the physical properties of a
  particular static naked singularity toy model which is supported by
  tangential pressure. The aim is to study differences between black
  hole and naked singularity configurations and to understand
  observational signatures that might be used to distinguish naked
  singularities and black holes of the same mass.  We focus on the
nature of stable circular orbits in a chosen metric and consider the
properties of accretion disks.

In the following, we discuss a specific model where we choose the mass
function to be $F(r)=M_0r^3$ such that the regularity conditions are
fulfilled during collapse. The divergence of the energy density in the
limit of the equilibrium configuration is then achieved by a choice of
$v_e(r)$ such that $v_e(0)=0$. As an example, we consider
$v_e(r) = cr^\alpha$ where, for simplicity, we set $c=1$ (thus
imposing a scaling in the boundary conditions).
 It is easy to verify that the value
$\alpha=0$ corresponds to a regular solution with positive density,
namely the `constant density' interiors first studied by Florides.
On the other hand,
$\alpha>0$ gives $v_e(0)=0$ and implies
the presence of a naked singularity at $r=0$ in equilibrium.
These singular interior models differ from other
regular interiors for Schwarzschild in the behaviour of
the density and curvatures near the center
(see for example
\cite{interiors}
for other regular interior solutions with perfect fluid
sources and/or cosmological constant).
The choice of $v_e(r)$ determines
the mapping between the physical radius R and
the comoving coordinate $r$ in the static metric,
$R(r)=r v_e(r)=r^{\alpha+1}$, and this, together with
the choice of $F(r)$, fixes the static solution.
In the specific toy model considered here,
we have
\begin{equation}
F(r)=M_0r^3, \quad v_e(r)=r^\alpha, \quad F(R)=M_0R^\frac{3}{\alpha+1}.
\label{toymodel}
\end{equation}

From the Einstein equations we obtain expressions for the equilibrium
density $\rho_e$ and pressure $p_{\theta e}$,
\begin{eqnarray}
  \rho_e &=& \frac{3M_0}{(\alpha+1)}\frac{1}{R^{\frac{3\alpha}{\alpha+1}}} \; ,\\
  p_{\theta e} &=& \frac{3M_0^2}{4(\alpha
    +1)}\frac{R^{\frac{2-4\alpha}
{\alpha+1}}}{\left(1-M_0R^{\frac{2-\alpha}{\alpha+1}}\right)} \; .
\end{eqnarray}
Different values of $\alpha$, which correspond to different choices of
$F(R)$, lead to different behaviours for the pressure as $r\rightarrow
0$.  We see that $\alpha<1/2$ implies $p_{\theta e}\rightarrow
0$, while $\alpha>1/2$ implies $p_{\theta e}\rightarrow
+\infty$; the transition value $\alpha= 1/2$ implies
$p_{\theta e}\rightarrow {\rm const}$.

In general, to understand the properties of accretion disks in the
static tangential pressure spacetimes as given by
equation \eqref{stat}, let us
consider test particles in circular
orbits. Without loss of generality, we take the orbits to be in the
equatorial plane ($\theta=\pi/2$).  Since the static metric is
independent of $t$ and $\phi$, we have two conserved quantities, the
energy per unit mass, $E= u_t = e^{2\nu} (dt/d\tau)$, and the
angular momentum per unit mass, $\ell = u_\phi = R^2
(d\phi/d\tau)$.  The normalization condition $u^\alpha
u_\alpha=-1$ then gives
\begin{equation}
\frac{1}{G}\left(\frac{dR}{d\tau}\right)^2-E^2e^{-2\nu}
+\left(1+\frac{\ell^2}{R^2}\right)=0 \; . \label{circ}
\end{equation}
For circular orbits, we set $dR/d\tau=0$, so we require the remaining
terms in the above equation to add up to zero. In addition, their sum
should achieve an extremum at radius $R$.  This gives the two
conditions
\begin{eqnarray}\label{E}
  E^2 &=& 2e^{2\nu}\left(\frac{R-F}{2R-3F}\right) \; , \\ \label{l}
  \frac{\ell^2}{R_b^2} &=& \frac{F}{2R-3F}
\left(\frac{R}{R_b}\right)^2 \; ,
\end{eqnarray}
where $R_b$ is the physical radius corresponding to the boundary of
the matter cloud in the final equilibrium state.
 It is to be noted that since we are
considering accretion disks which rotate freely in a metric
that describes an internal fluid, we have to assume
that the fluid constituting the naked singularity is
weakly interacting with the matter
constituting the accretion disk, so that the particles in
the disk can have circular geodesic motion.

From equations \eqref{E} and
\eqref{l} we find that,
for $R_b<3\mathbf{M_{TOT}}$, both the quantities
$E^2$ and $\ell^2$ become negative, thus indicating that the
accretion disk particles must have imaginary
energy and angular momentum to move on circular geodesics.
This result is true also
for perfect fluid interiors describing static sources of
the Schwarzschild spacetime. Furthermore, for $R<2.5\mathbf{M_{TOT}}$,
the sound speed within the cloud becomes superluminal, which is
unphysical. For all these reasons, in the following we focus on
models with $R_b>3\mathbf{M_{TOT}}$.

In order to understand the properties of these
naked singularity models better and to compare them with
the Schwarzschild black hole case, we now consider
a specific example, viz.,
models with $\alpha=2$. In this case, both the energy
density and the pressure diverge at the
center as $R^{-2}$. From the Misner-Sharp mass, we see that at the boundary
$2\mathbf{M_{TOT}}/R_b = M_0$. Thus the energy
density is given by $\rho_e = M_0/R^2$ and the pressure
satisfies a linear equation of state, $p_{\theta e}=k\rho_e$, with
$4k=M_0/(1-M_0)$.  In this simple model, the condition to
avoid an event horizon is specifically $M_0<1$.  Furthermore,
to satisfy the weak energy condition, we must have $k\geq -1$,
which corresponds to $M_0\leq 4/3$. The effective sound
speed $c_\theta$ is given by $c_\theta^2 = p_{\theta e}/\rho_e
= k$, and if we want this to be less than unity we then
require $M_0 < 4/5$.

From the Einstein equations we find $2\nu(R) =
\ln\left[CR^{M_0/(1-M_0)}\right]$, where $C$ is an integration
constant that can be evaluated from the boundary condition.
Thus we obtain
\begin{equation}
e^{2\nu(R)} =
(1-M_0)\left(\frac{R}{R_b}\right)^{M_0/(1-M_0)}.
\end{equation}
The complete solution for the metric in the interior
$R<R_b$ is then given by,
\begin{equation}
ds^2_e =  -(1-M_0)\left(\frac{R}{R_b}\right)^{\frac{M_0}{1-M_0}}dt^2+
\frac{dR^2}{1-M_0}+R^2d\Omega^2 \; .
\end{equation}
This metric matches smoothly to a Schwarzschild spacetime
in the exterior $R\geq R_b$,
\begin{equation}
ds^2= -\left(1- \frac{M_0 R_b}{R}\right)dt^2+\frac{dR^2}
{(1- M_0 R_b/R)}+R^2d\Omega^2 \; .
\end{equation}
We thus have a one-parameter family of static equilibrium solutions
parametrized by $M_0$ (in principle, there is a second parameter $R_b$,
but this is simply a scale). Each member of this family of solutions has a naked singularity
at the center. As described earlier, these solutions can be
obtained as the end state of dynamical collapse from regular
initial conditions with $F(r)=M_0r^3$, by choosing the evolution
function $v(r,t)$ such that it asymptotes to the required
$v_e(r) \propto r^2$ as $t\to\infty$ (see equation \eqref{toymodel}).

In order to specify the nature of the central singularity,
we note that the outgoing radial null geodesics in the spacetime
above are given by,
\begin{equation}
\frac{dR}{dt}=(1-M_0)\left(\frac{R}{R_b}\right)^{\frac{M_0}{2(1-M_0)}} \; .
\end{equation}
It is then easy to check that there are light rays
escaping from the singularity (for all values of $M_0<2/3$).
In fact from the above equation which gives,
\begin{equation}
    t(R)=\frac{2R_b^{\frac{M_0}{2(1-M_0)}}}{2-3M_0}R^{\frac{2-3M_0}{2(1-M_0)}} \; ,
\end{equation}
we see immediately that the comoving time required by a photon
to reach the boundary is $t_b=2R_b/(2-3M_0)<+\infty$.
It follows that there are future directed null geodesics
in the spacetime which reach the boundary of the cloud, and which
in the past terminate at the singularity, thus showing this
to be a naked singularity. The density and spacetime curvatures
would diverge in the limit of approach to the singularity in
the past along these null trajectories, showing this to be
a curvature singularity.
The Kretschmann scalar for this naked
singularity model, for the $\alpha=2$ case, is given by
\begin{equation}
    K = \frac{1}{4} \frac{M_0^2(28-60M_0+33M_0^2)}{(M_0-1)^2 R^4} \; .
\end{equation}
We see that the Kretschmann scalar diverges in the
limit of approach to the central singularity.
The spacetime is regular everywhere for all values of $r>0$.
A similar situation holds also for other models in this class
of tangential pressure solutions.
The divergence of curvatures at the center clarifies
that this is a genuine spacetime singularity. Then the
absence or otherwise of a trapped surface would determine
whether this is a naked singularity or not, which again
it is in this case, as we have shown above.

Circular geodesics for accretion disks in singular spacetimes
without an event horizon have been studied in a variety of
scenarios that include static and stationary spacetimes with
and without a scalar field
(\cite{Kerr} and see also \cite{HR}).
The motion of test particles in circular orbits in a given
spacetime is characterized by the existence of certain key
parameters such as the photon sphere, the minimum radius
for bound circular orbits and the minimum radius for stable
circular orbits.
As we shall see below the main features that stand out for our
static toy model are the absence of an
innermost stable circular orbit, meaning that stable orbits
extend all the way to the singularity, and absence of
the photon sphere. This marks a sharp contrast with similar
analysis in some other naked singular static and stationary
geometries where the presence of both a minimum radius for stable
orbits and a photon sphere make the objects virtually indistinguishable
from a black hole, at least as far as their optical properties
are concerned
(see \cite{Virb}).

For the toy naked singularity model under consideration, the energy
per unit mass $E$ and
angular momentum per unit mass $\ell$ of the circular orbits
may be obtained from equations \eqref{E} and \eqref{l}.
For $R < R_b$, we find,
\begin{eqnarray}\label{E2}
  E^2 &=& \frac{2(1-M_0)^2}{(2-3M_0)}\left(\frac{R}{R_b}
\right)^{M_0/(1-M_0)} \; ,\\ \label{L2}
  \frac{\ell^2}{R_b^2} &=& \frac{M_0}{(2-3M_0)}\left(\frac{R}{R_b}\right)^2 \; .
\end{eqnarray}
If we want the circular orbit calculated above to be stable, we
require in equation (\ref{circ}) that the term involving $E^2$ should
be less divergent as $R\to0$ compared to the term involving $\ell^2$.
This then gives the following results for $R<R_b$:
\begin{eqnarray}
{\rm Stable~circular~orbits:}&~&\qquad M_0\leq 2/3, \\
{\rm Unstable~circular~orbits:}&~&\qquad M_0> 2/3.
\end{eqnarray}
We see that, depending on the value of $M_0$, either all circular
orbits in the interior of this naked singularity model are stable,
or all are unstable. Note that, apart from having unstable
circular orbits, models with $M_0>2/3$ also give
negative values of $E^2$ and $\ell^2$.

For $R\geq R_b$, the metric is given by the Schwarzschild solution
with mass $\mathbf{M_{TOT}}=M_0R_b/2$. Here we have
well-known results for the stability of circular orbits, viz.,
orbits with $R\geq 6\mathbf{M_{TOT}}$ are stable, while
those with $R<6\mathbf{M_{TOT}}$ are unstable.
Further, the space-time has closed circular photon orbits at
$R=3\mathbf{M_{TOT}}$ (assuming this radius is located
outside $R_b$).

The practical significance of the above results is related
to the fact that a standard thin accretion disk can exist only
at those radii where stable circular orbits are available
\cite{NovikovThorne}.
Thus, for a Schwarzschild black hole of mass $\mathbf{M_{TOT}}$, an
accretion disk will have its inner edge at the innermost stable
circular orbit at $R=6\mathbf{M_{TOT}}$. Inside this radius,
the gas plunges or free-falls until it crosses the horizon.
The existence of a well-defined disk inner edge
\cite{diskedge},
which is the basis for the
well-known Novikov-Thorne model of a relativistic thin accretion disk
around a black hole
\cite{NovikovThorne},
will be reflected in the radiation spectrum emitted by the disk.
Indeed, observations have confirmed the presence of such an edge
in several cases
\cite{Steiner}.
Moreover, assuming that the central object is a black hole,
the radius of the disk inner edge has been used to estimate
the spin parameters of the black holes
\cite{BHspin}.

It is worth noting from equations
\eqref{E2} and \eqref{L2} that both the energy per unit mass $E$
and the angular momentum per unit mass $\ell$
of the gas in the accretion disk vanish in the limit of $R=0$.
This means that no mass or rotation is added to the central
singularity by the accretion disk,
whose particles radiate away or otherwise get rid of all their energy and angular
momentum before reaching the singularity.
It thus follows that the process of accretion does
not affect the naked singularity, which can be considered
stable in this respect.
This is very different from what is expected to happen in
the case of a rotating Kerr naked
singularity, where the process of accretion of counter-rotating
particles can spin down
the object and turn it into a near-extremal black hole
(see for example
\cite{Stuch}).

We considered above several general physical features
of the toy naked singularity model, many of
which also apply to the general class of static tangential
pressure models with a naked singularity.
In general, the observational
properties of accretion disks can be characterized
in terms of the energy flux, luminosity, and also the spectrum
of the emitted radiation.
Such features have been analyzed recently within certain models
with naked singularities which are present in extremal
stationary Kerr spacetime geometries
(see e.g. \cite{HR}).
Such studies might possibly help towards
observationally distinguishing black holes from
naked singularities. From such an investigation of different
aspects of accretion disks we may be able to reveal
the crucial features that make them different from the
widely studied accretion disks in the Kerr spacetime.
It has to be noted, however, that many of the models
with static or stationary naked singularities, such as the
Janis-Newman-Winicour (JNW) spacetime, Reissner-Nordstrom geometries with
$Q > M$ or superspinning Kerr geometries need not arise
naturally from dynamical evolution in gravitational collapse
under Einstein's equations.
Our model, on the other hand, provides a dynamical
framework through which a static compact object with a naked
singularity at its core can be formed as the limit of gravitational
collapse of a massive matter cloud with non-zero
tangential pressure.

A detailed analysis of the properties of accretion disks
for interior solutions with
tangential pressure, both in the regular and singular cases,
is beyond the scope of this paper and
will be discussed in detail elsewhere.
Our main purpose here is a comparison of these naked singularity
objects which can form via collapse, with a Schwarzschild
black hole of the same mass.
Therefore, making use of the stability properties of
circular orbits, we identify the following two distinct
model regimes, each with its own accretion structure, as we
discuss below:

\noindent
$\bullet$ $M_0\leq1/3$, i.e., $R_b\geq 6\mathbf{M_{TOT}}$:\\ In this
case, the external Schwarzschild metric has stable circular orbits all
the way down to the boundary $R=R_b$ where it meets the interior
metric of our naked singularity solution. Consequently, an
accretion disk will follow
the standard Novikov-Thorne disk solution down to $R=R_b$. Inside
$R_b$, the interior metric allows stable circular orbits all the way
down to $R=0$. Thus, the disk will continue into the interior and will
extend down to $R=0$. In other words, the disk will have no inner
edge. Assuming the matter cloud that makes up the naked singularity is
transparent to radiation (we have already assumed that it does not interact
with the gas in the accretion disk), a distant observer will receive radiation
from all radii down to the center and the observed spectrum will obviously be
very different from that seen from a disk around a black hole of the
same mass.  (We postpone detailed computation of the spectrum to a
later investigation.) As an aside, note that this space-time has no
 circular null geodesics (photon sphere).\\

\noindent
$\bullet$ $1/3<M_0\leq2/3$, i.e., $6\mathbf{M_{TOT}} > R_b\geq
3\mathbf{M_{TOT}}$:\\ In this case, an accretion disk will follow the
Novikov-Thorne solution down to $R=6\mathbf{M_{TOT}}$.  Inside this
radius, since circular orbits are unstable in the Schwarzschild
space-time, the gas will plunge towards smaller radii. However, once
the gas reaches the boundary of the interior solution at $R=R_b$,
circular orbits are once again available.  Hence, we expect the gas to
shock and circularize at $R=R_b$ and then to continue
accreting along a sequence of stable circular orbits
all the way down to $R=0$.  (We
assume that the gas at $R_b$ can get rid of its excess angular
momentum by some means to the outer disk across the gap.)  Since the
accretion disk in this model consists of two distinct segments with a
radial gap in between, we expect it to be observationally
distinguishable from the previous case.  Once again, there is no
photon sphere in this space-time.

As we have mentioned before, the  models with
$3\mathbf{M_{TOT}} > R_b>2\mathbf{M_{TOT}}$ (corresponding to
  $2/3<M_0<1$) present unphysical and exotic properties
that would indicate that the boundary of the final
static configuration
should be taken at a value larger than $3\mathbf{M_{TOT}}$.
We note that for the accretion regimes we considered above,
which is the range as given by $M_0 < 2/3$, all reasonable
physical properties for the matter fields are satisfied by
the particles of the accretion disk.

The interesting point is that due to
the absence of a photon sphere, naked singularity models with $M_0<2/3$ are easily
distinguishable from a black hole of the same mass.
This opens up the possibility of using observational data on
astrophysical black hole candidates to test for the presence
of a naked singularity. The discussion here pertains only
to the particular toy model with $\alpha=2$. Models with other
values of $\alpha$, or more generally, models in which $v_e(r)$
is more complicated than a power-law in $r$, may well give other
kinds of behavior that may be worth investigating.

\begin{figure}[hh]
\includegraphics[scale=0.8]{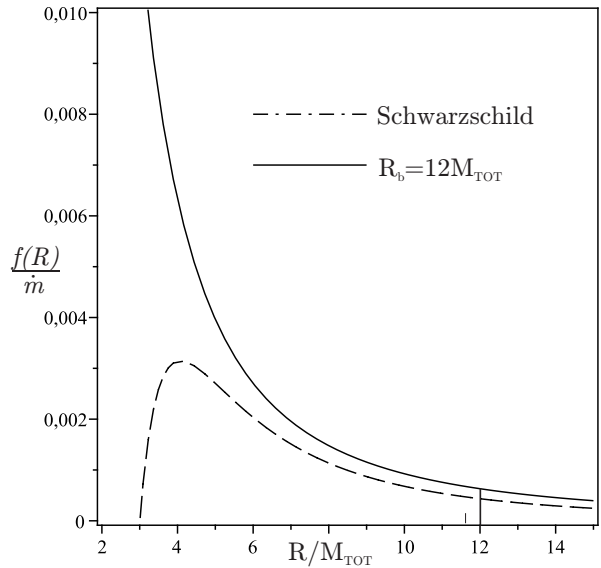}
\caption{Comparison between the radiant energy flux
for an accretion disk around a Schwarzschild black hole with
total mass $\mathbf{M_{TOT}}$ (dashed line),
and for an accretion disk in the toy naked singular
model with the same mass and $R_b=12\mathbf{M_{TOT}}$
(solid line).}
\label{fig3}
\end{figure}

A key quantity in the case of an accretion disk
is the radiant energy flux as a function of radius. This is given by
\begin{equation}\label{flux}
    f(R)=-\frac{\dot{m}}{\sqrt{-g}}\frac{\omega_{,R}}
{(E-\omega \ell)^2}\int^R_{R_{in}}(E-\omega \ell)\ell_{,R}dR \; ,
\end{equation}
where $R_{in}$ is the radius of the inner edge of
the disk,
$\dot{m}$ is the mass accretion rate, which for steady state
accretion is usually assumed to be constant, and
$\omega=d\phi/dt$ is the angular velocity of
particles on circular orbits.
For a Schwarzschild black hole, $R_{in}=6\mathbf{M_{TOT}}$
and $f(R)$ vanishes for smaller radii. This leads to well-known
results that are widely used for
modeling accretion disks observed in astrophysics. In the
case of our naked singularity model, the inner edge of the disk
is at $R=0$. Therefore, the flux continues to increase with
decreasing $R$, diverging at $R=0$ (the integrated total
luminosity observed at infinity is of course finite).
The disk ends up being
much more luminous as compared to the black hole case
(see Fig. \ref{fig3}) and the two cases may be easily distinguished.
The fact that $f(R)$ diverges as $R$ goes
to zero is not surprising and is related to the presence of the singularity.

Although we decided earlier to focus on models with $M_0<2/3$,
for completeness we briefly discuss here the parameter range
$2/3<M_0<1$. For a model in this range,
an accretion disk will follow the
Novikov-Thorne solution down to $R=6\mathbf{M_{TOT}}$. The lack of
stable circular orbits inside this radius will
then cause the gas to plunge inwards.
When the gas crosses $R=R_b$, there are still no stable
circular orbits available (in contrast to the previous cases),
so the gas will continue to plunge all the
way down to $R=0$.  The accretion disk in this case is easily
distinguishable from each of the previous two cases.  However,
for various reasons, it will
most likely be indistinguishable from a standard
Novikov-Thorne disk around a Schwarzschild black hole of
mass $\mathbf{M_{TOT}}$. Firstly, although
the fate of the gas that reaches the naked singularity at the center
is unclear, since this gas carries energy and angular momentum it will
most likely modify the nature of the central singularity.
Secondly, any radiation
that is emitted from the singularity will not escape to infinity,
as we showed earlier. Both arguments suggest that this model will
behave for all practical purposes like a black hole.
This regime is further divided into two subregimes since
for $4/5 \le M_0 < 1$
we showed that the sound speed exceeds unity.
Note that there is a photon sphere at the standard
Schwarzschild location, $R_{\rm photon}=3\mathbf{M_{TOT}}$.

All of the above discussion pertains to a
model with $\alpha=2$. It is interesting to note here, for
the sake of a comparison, that in a model with $\alpha=0$,
which corresponds to
a regular static solution with non-zero tangential pressure,
there are again two different regimes according
to where we take the boundary of the cloud.
The first regime corresponds to $R_b\geq 6\mathbf{M_{TOT}}$,
and in this case there
is no photon sphere and stable circular orbits extend all the way down to
the regular center. In the second regime,
$6\mathbf{M_{TOT}}>R_b\geq 3\mathbf{M_{TOT}}$, the matter
in the accretion disk reaches the last stable circular orbit of
the Schwarzschild spacetime at $R=6\mathbf{M_{TOT}}$,
then plunges down to the boundary of the interior solution $R=R_b$,
inside which stable orbits
are again allowed down to the center. Again there is no photon
sphere. The situation in this respect is thus similar to the
first two cases studied above for the naked singularity model with $\alpha=2$.
The main difference for the $\alpha=0$ case is that particles
in the accretion disk reach the regular center with non-vanishing
energy (and vanishing angular momentum). It is useful to note
that again, for
$R_b<3\mathbf{M_{TOT}}$ the energy and angular momentum have to
be imaginary if a particle is to follow a circular geodesic,
and for $R_b<(5/2)\mathbf{M_{TOT}}$ the effective
sound speed surpasses unity while close to
the boundary.

The naked singularity model presented here, which arises from the
dynamical gravitational collapse of a massive matter cloud
with non-zero tangential pressure, presents several interesting
physical features some of which we have analyzed here.
In particular, the accretion disk properties
allow it to be distinguished observationally from
a Schwarzschild black hole with the same mass. There are several
other physical properties which are worth studying. Particular
mention should be made to optical phenomena, where
the toy naked singularity model discussed here will have
quite different behavior compared to
Kerr and certain
other naked singularity spacetimes.
The main important difference, as far as optical
properties and gravitational lensing are concerned, is the
following. All the other naked singularity models
mentioned above and discussed earlier necessarily admit the
presence of a photon sphere for a certain range of the
solution parameters involved.
For example, the Reissner-Nordstrom
spacetime with $Q > M$, which has a naked singularity, or the
JNW naked singularity, necessarily admit a photon
sphere when the quantities $Q-M$ and the scalar charge $\mu$ are
respectively small enough. Such naked singularities have been
termed as `weakly naked singular'
(see e.g. \cite{Virb}), and
are expected to be observationally indistinguishable
from black holes, especially as far as their optical properties
are concerned.
As opposed to this, the tangential pressure naked singularity
models presented here have no photon sphere as discussed
above. Therefore these are necessarily `strongly naked singular'
and will be distinguishable always from black holes.

\section{Concluding remarks}\label{remarks}

We have investigated here the equilibrium
configurations that can be achieved from gravitational
collapse of a spherical matter cloud with vanishing
radial pressure. We showed that all static interiors
of the Schwarzschild space-time with tangential pressure can be
obtained as the limit of some model for dynamical
gravitational collapse. These static interiors might be
regular or they may have a naked singularity at the center.

The key important features of
this model that distinguish it from
other naked singular spacetimes are:
\begin{itemize}
  \item The naked singularity is obtained via dynamical
  evolution of a matter cloud starting with regular initial data.
  \item For the particular class of toy models with $\alpha=2$,
    $M_0<2/3$, that we have focused on, the singularity is not
    destroyed by the infall of particles through an accretion disk and
    hence it is stable in this sense.
  \item Due to the absence of a photon sphere in these solutions, the
    singular spacetime is always optically distinguishable from a
    black hole with the same mass.
\end{itemize}

We also examined and noted here several
physical properties and features of accretion disks in
these naked singularity models, comparing them with those
for a Schwarzschild black hole, and we noted how black
holes and naked singularities will have observationally
distinct signatures (e.g., see Fig.~\ref{fig3}).

In analogy with the Newtonian case, although
the equilibrium configurations we describe can be reached via
a wide class of pressure evolutions, they sit at the maximum
of the effective potential (see Fig.~\ref{fig}) and are
expected to be unstable under small perturbations
in the velocities.
Therefore, tangential pressure models $p_\theta(r,v)$
close to the ones leading to an equilibrium, but with
a different asymptotic behaviour, will lead to either
complete collapse or rebounce. Nevertheless, the main
point we wish to make is that static equilibrium
configurations as a limit to gravitational collapse do
arise and exist, and that the formalism for collapse in general
relativity does not always imply that the matter cloud must
necessarily collapse under its own gravity to a final
singularity in a `short' time. In fact, since the
equilibrium configurations described here are reached
only in the limit of $t$ going to infinity, all
neighbouring solutions (meaning those tangential
pressure evolutions that have an asymptotic limit close
to equilibrium) can be `long lived' and could describe
systems that evolve over an arbitrarily long time. In
this sense, the equilibrium configurations investigated
here constitute a valid toy model to describe
`long lived' dynamical models, where the collapse
essentially `freezes' as it evolves in time.

We investigated in section IV one
specific static equilibrium solution ($\alpha=2$) with
a naked singularity at the center, and we showed
that the accretion properties of such an object
can, in principle, be quite different from those of
a Schwarzschild black hole.
Other models with different values of the parameter
$\alpha$, e.g., the Florides
constant density interior solution ($\alpha=0$),
or a different functional behaviour of $v_e(r)$,
e.g., $v_e \simeq cr^\alpha$
only near $r=0$ and having a different radial variation
away from the center, could be investigated as well.
Other physical
features such as gravitational lensing or the
properties of the photon sphere could also be considered
in more detail in order to have a better understanding
of the physical nature and properties of these
theoretical models.

Recently there has been some interest in the possibility of
observationally distinguishing black holes from naked singularities.
Most of these studies deal with naked singularities
that are present in extremal stationary Kerr spacetimes and therefore
need not arise naturally from dynamical evolution under
Einstein's equations. Our model, on the other hand, provides a dynamical
framework through which a compact object with a naked singularity
at its core can be formed.

The formalism developed in this paper might be applied
in an astrophysical context to describe
the final fate of gravitating objects collapsing
under the force of their own gravity.
There have been detailed investigations of the last
stages of evolution of a massive star when all the radial shells of matter
fall towards a central singularity to make a black hole.  From the
considerations described here other end-states
are also possible, e.g. the system could asymptote to a
static solution with or without a naked singularity.
The class of static singular solutions might conceivably
be of use to describe rare astrophysical phenomena.

\begin{acknowledgments}
    PSJ would like to thank Peter Biermann 
    for discussions on the object-like and event-like character 
    of singularities when they are visible.
\end{acknowledgments}

\end{document}